\documentclass[nofootinbib,superscriptaddress,a4paper,onecolumn]{revtex4-2}
\usepackage{geometry}
\usepackage{array,mathtools,amssymb,booktabs}
\usepackage{amsmath, amsfonts, mathrsfs}
\usepackage{hyperref}

\usepackage{graphicx}

\newcommand{\mic}{\textmu{}m~}

\begin{document}

\title{Broadband behavior of quadratic metalenses with a wide field of view}

\author{Yang Liu}
 \altaffiliation[Currently with ]{DTU Fotonik, Technical University of Denmark, Ørsteds Plads 343, 2800 Kgs. Lyngby, Denmark.}%
 \affiliation{Centre de Nanosciences et de Nanotechnologies, Université Paris-Saclay, CNRS, 91120 Palaiseau, France}
\author{Jianhao Zhang}%
\author{Xavier Le Roux}%
\author{Eric Cassan}%
\author{Delphine Marris-Morini}%
\author{Laurent Vivien}%
\author{Carlos Alonso-Ramos}%
\author{Daniele Melati}%
\email{daniele.melati@universite-paris-saclay.fr}
\affiliation{Centre de Nanosciences et de Nanotechnologies, Université Paris-Saclay, CNRS, 91120 Palaiseau, France}

%%%%%%%%%%%%%%%%%%% abstract %%%%%%%%%%%%%%%%
\begin{abstract}
Metalenses are attracting a large interest for the implementation of complex optical functionalities in planar and compact devices. However, chromatic and off-axis aberrations remain standing challenges. Here, we experimentally investigate the broadband behavior of metalenses based on quadratic phase profiles. We show that these metalenses do not only guarantee an arbitrarily large field of view but are also inherently tolerant to longitudinal and transverse chromatic aberrations. As such, we demonstrate a single-layer, silicon metalens with a field of view of 86° and a bandwidth up to 140 nm operating at both 1300 nm and 1550 nm telecommunication wavelength bands.
\end{abstract}

\maketitle

%%%%%%%%%%%%%%%%%%%%%%%%%%  body  %%%%%%%%%%%%%%%%%%%%%%%%%%
\section{Introduction}
Metasurfaces are two-dimensional arrangements of subwavelength scattering elements capable of manipulating the phase, amplitude, and polarization of a wave front \cite{lalanne_design_1999, yu_light_2011}. This technology has been successfully exploited to realize compact and flat planar devices implementing a variety of optical functionalities, including lenses, holograms, mirrors, and polarizers, to name a few \cite{joo_metasurface-driven_2020, kamali_review_2018, kildishev_planar_2013}.
The possibility to precisely tailor the optical response of each element in the metasurface (meta-atom) provides an invaluable design freedom. This notoriously enabled, among other examples, the demonstration of high-performance metalenses free of spherical aberration for on-axis propagation reaching diffraction-limited resolution at a specific wavelength \cite{aieta_aberration-free_2012, khorasaninejad_metalenses_2016}.
However, despite great findings, simultaneously ensuring broadband operation and wide field of view still remains challenging for metalenses \cite{presutti_focusing_2020, lalanne_metalenses_2017,liang_high_2019}, an aspect of fundamental importance in many applications including imaging and beam steering.
A number of designs with compensation of chromatic aberration have been proposed, mostly relying on the judicious engineering of meta-atoms dispersion \cite{khorasaninejad_achromatic_2017, arbabi_controlling_2017, shrestha_broadband_2018, wang_broadband_2018, ndao_octave_2020}, but commonly restricted to on-axis propagation or small incidence angles.
Likewise, metalenses with a large field of view have been obtained using cascaded, multi-layer structures (doublets) \cite{arbabi_miniature_2016, groever_meta-lens_2017} and aperture stops \cite{shalaginov_single-element_2020}, but limited to monochromatic operation or moderate bandwidths \cite{engelberg_near-ir_2020}.
Very recently, achromatic metalenses demonstrating also a wide filed of view have been achieved using optimized phase profiles designed via deep learning techniques \cite{yang_design_2021} and exploiting metalens doublets \cite{huang_achromatic_2021}.

Interestingly, it has also been shown that it is possible to achieve an arbitrarily wide field of view trading-off the need for diffraction-limited resolution and designing metalenses with a quadratic phase profile, which provides a point spread function that remains undistorted upon tilted illumination   \cite{pu_nanoapertures_2017,guo_high-efficiency_2018, martins_metalenses_2020,lassalle_imaging_2021}. While the single-wavelength performance of such metalenses has been described in details, to our knowledge their broadband behaviour has not yet been extensively studied, limiting their potential applicability in multi-wavelength systems. Only an initial qualitative demonstration of a rather large tolerance to longitudinal chromatic aberration has been reported \cite{pu_nanoapertures_2017}.
In this work, we experimentally investigate, for the first time, the broadband behavior of a single-layer, quadratic metalens over two wavelength ranges at typical the telecommunication wavelengths of 1300 nm and 1550 nm. We show that, due to the large depth of focus, the size of the lens focal spot remains unchanged across both wavelength bands, quantitatively confirming the tolerance to longitudinal chromatic aberration. Furthermore, we also demonstrate that the reduced resolution of quadratic metalenses helps in achieving a large tolerance to transverse chromatic aberration, an aspect that has not been addressed before but of central importance for metalenses with a wide field of view. Overall, the metalens, realized on a silicon platform and with a nominal numerical aperture of 0.83, achieves simultaneously a measured spectral bandwidth of 140 nm and a wide field of view of 86°, limited only by our experimental setup.

\section{Quadratic metalens design}
\begin{figure}[t]
\centering\includegraphics[width=0.98\textwidth]{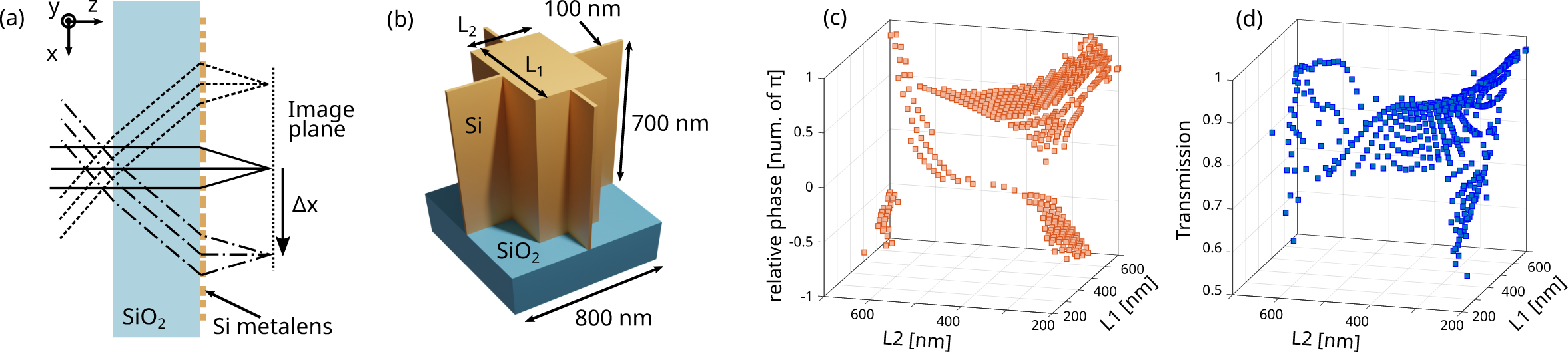}
\caption{Design of the quadratic metalens. (a) Schematic diagram of the metalens operation. The shape of the focal spot does not change with oblique incidence but only shift on the image plane. (b) Schematic of the meta-atom used to realize the fishnet metalens. The two dimensions of the silicon pillars L$_1$ and L$_2$ are independently chosen. (c) Phase and (d) amplitude response of the meta-atoms as a function of L$_1$ and L$_2$.}
\label{fig: design}
\end{figure}
The phase delay imparted by a quadratic metalens to an incoming wavefront is described by
\begin{equation}
\label{eq: phase_profile}
\varphi(r) = -\frac{\omega n_{\rm f}}{2cf}r^2.
\end{equation}
In Eq. \eqref{eq: phase_profile}, $\omega$  is the frequency of the light, $n_{\rm f}$ is the refractive index of the material where light is focused, $c$ is the speed of light, $f$ is the focal distance of the metalens, and $r$ is the radial distance from the center of the lens. Assuming the metalens lies in the xy plane and a plane wave arrives with an incident angle $\theta$ from the normal in the xz plane, propagating in a medium with refractive index $n_{\rm c}$, the phase front after the metalens becomes
\begin{equation}
\label{eq: phase_shift}
\varphi(r) = -\frac{\omega n_{\rm f}}{2cf}r^2 - \frac{\omega}{c} n_{\rm c} x\sin{\theta} = 
-\frac{\omega n_{\rm f}}{2cf}\left[\left(x + \frac{n_{\rm c}}{n_{\rm f}}f\sin{\theta}\right)^2 + y^2\right] + \frac{\omega f  n_{\rm c}^2 \sin^2{\theta}}{2cn_{\rm f}}
\end{equation}
The second term in Eq. \eqref{eq: phase_shift} does not depend on the radial distance and can be neglected since it does not influence the shape of the wavefront. Comparing the first term of Eq. \eqref{eq: phase_shift} with Eq. \eqref{eq: phase_profile}, it can be seen that the oblique incidence results in a transverse shift of the focal spot in the x direction, as schematically illustrated also in Fig. \ref{fig: design}(a):
\begin{equation}
\label{eq: transversal_shift}
\Delta x = \frac{n_{\rm c}}{n_{\rm f}}f\sin{\theta}
\end{equation}
The use of a quadratic phase profile hence allows to achieve a wide field of view by focusing light beams incident at different angles on a planar image plane with an undistorted focal spot. However, it should be noticed that the quadratic phase profile also introduces spherical aberration and a large depth of focus \cite{pu_nanoapertures_2017}. This results in a reduced resolution compared for example to diffraction-limited metalenses based on an hyperbolic phase profile, as thoroughly discussed in \cite{martins_metalenses_2020}.

In order to implement the described phase profile we use silicon meta-atoms with a height of 700 nm on SiO$_2$, as shown in Fig. \ref{fig: design}(b). A fishnet structure is chosen to ensure a better mechanical stability of the meta-atoms. The metalens is realized with a lattice period of 800 nm that avoids high-order diffraction in the far field, provides a sufficiently accurate sampling of the phase profile, and enables a large feature size to ensure a reliable fabrication. To design the metasurface, the dimensions L$_1$ and L$_2$ of the rectangular pillars are varied between 200 nm and 700 nm. The cross width is fixed at 100 nm. 

For each combination of L$_1$ and L$_2$, we compute the phase and transmission of the meta-atom by simulating a corresponding periodic array for normal incidence through the Rigorous Coupled Wave Analysis \cite{hugonin2021reticolo}. We then disregard all meta-atoms with transmission efficiency lower than 0.5 to reduce the metalens losses and spurious amplitude modulations. Simulation results for $\lambda$ = 1550 nm are reported in Fig. \ref{fig: design}(c,d), respectively. The remaining library of meta-atoms allows to cover the entire phase range between $-\pi$ and $\pi$ and is eventually used to discretize the phase profile \eqref{eq: phase_profile} by properly organizing the meta-atoms according to the required phase $\varphi(r)$.

\section{Broadband behaviour of the metalens}
\label{sec: experiment}
\begin{figure}[t]
\centering\includegraphics[width=0.49\textwidth]{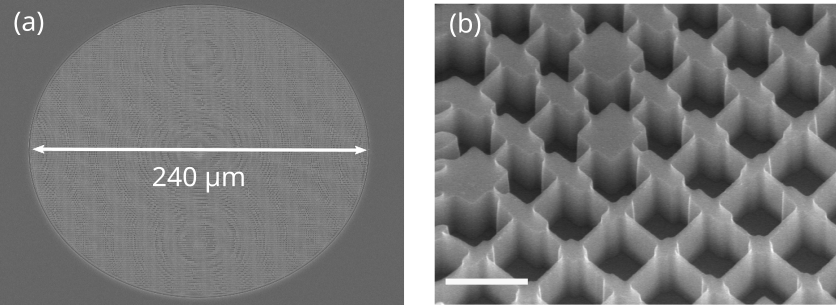}
\caption{Scanning electron microscope (SEM) pictures of the fabricated quadratic metalens. (a) Full device and (b) close-up on few lattice periods showing the quality of the fabricated device. The scale bar in (b) is 1 \mic.}
\label{fig: characterization}
\end{figure}
We fabricated a metalens with a diameter D = 240 \mic and a focal length in air $f$ = 80 \mic at a wavelength $\lambda$ = 1550 nm, resulting in a high nominal numerical aperture NA = 0.83. The fabrication was done using electron beam lithography and dry etching. Scanning electron microscope pictures of the realized device are shown in Fig. \ref{fig: characterization}. The collimated output of a tunable laser was used to illuminate the metalens at different angles and light wavelengths. The corresponding point spread function was then imaged on an InGaAs camera through a 40X objective with numerical aperture of 0.65 and a tube lens with focal distance of 200 mm. The metalens was mounted on a piezoelectric translation stage to allow imaging of the the three-dimensional intensity distribution of light after the metalens.
\begin{figure}[b]
\centering\includegraphics[width=0.98\textwidth]{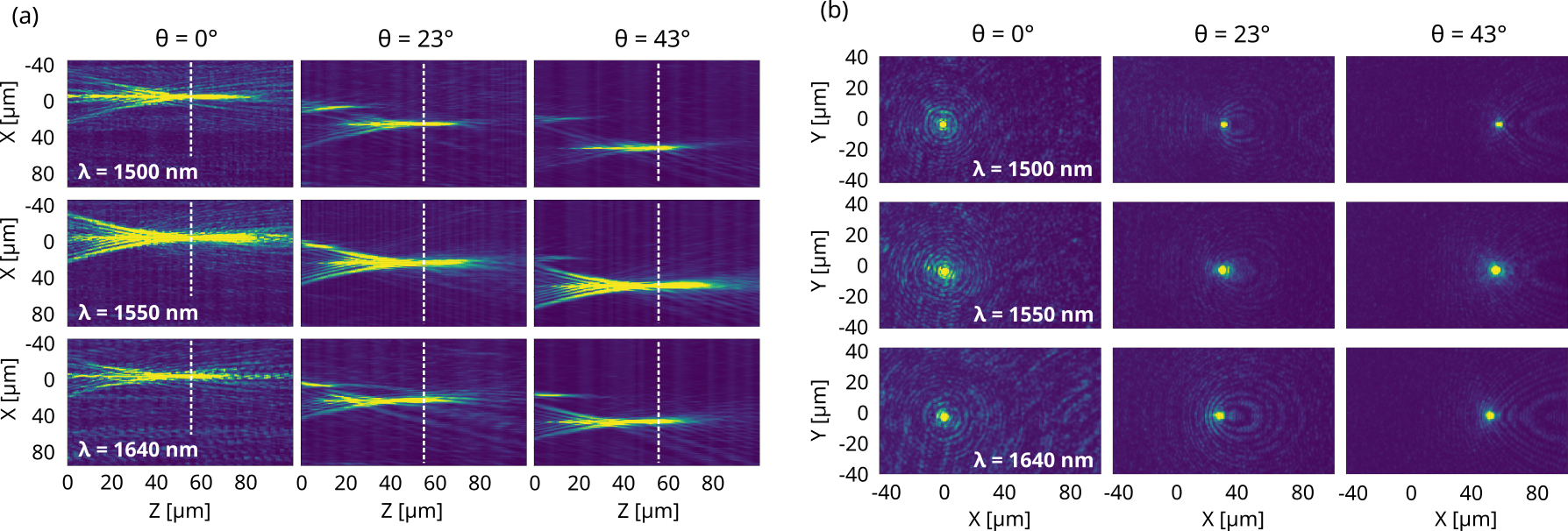}
\caption{Experimental results of broadband and wide-angle focusing. (a) Measured intensity profiles in the axial xz plane, being z the propagation direction, at different wavelengths from 1500 nm to 1640 nm and illuminating angles from 0° to 43°. (b) Intensity profiles at the image plane marked by the dashed lines in (a) (transverse xy plane) for the same wavelengths and angles.}
\label{fig: images_1}
\end{figure}

The measured intensity distributions in the horizontal axial plane (xz cross section, see Fig. \ref{fig: design}(a) for the reference axes) are plotted in Fig. \ref{fig: images_1}(a) for wavelength ranging from 1500 nm to 1640 nm and illumination angles from 0° to 43°. Our setup did not allowed to further increase the illumination angle. The origin of the z axis is arbitrary but fixed throughout the measurements. The corresponding image plane intensities (transverse xy plane) are shown in Fig. \ref{fig: images_1}(b). All focal spots are imaged at the same distance along the z axis, as marked by the white dashed lines in Fig. \ref{fig: images_1}(a). While longitudinal chromatic aberration causes the focal distance to reduce at longer wavelengths, the relative large depth of focus allows overcoming the effect of the shift and enables the metalens to operate at a constant image plane within a broad bandwidth. An almost unchanged focal spot is maintained across the entire wavelength range and also at different illumination angles. At tilted illumination, the position of the focal spot on the image plane simply laterally shifts along the x direction, as described by Eq. \eqref{eq: transversal_shift}. 

\begin{figure}[t]
\centering\includegraphics[width=0.49\textwidth]{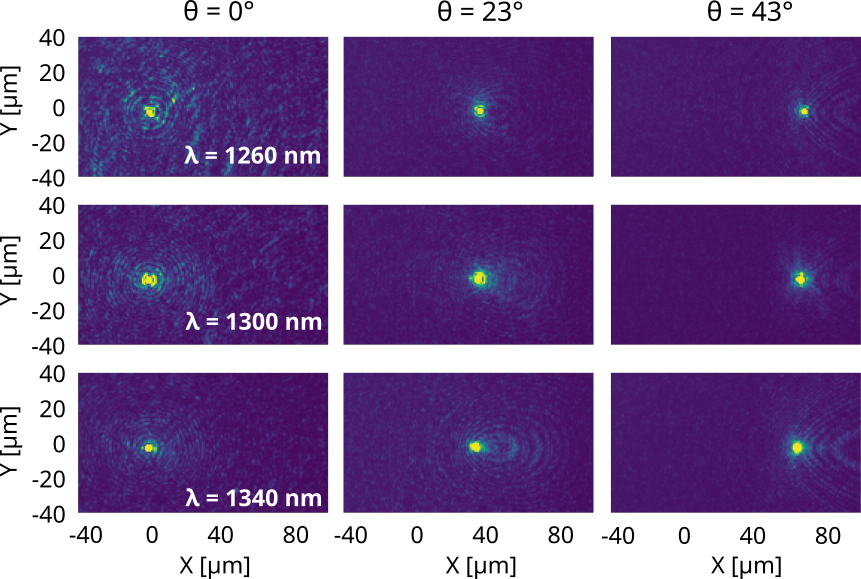}
\caption{Characterization in the 1300 nm wavelength range. Intensity profiles at the image plane (transverse xy plane) for the same illuminating angles of Fig. \ref{fig: images_1} but with wavelength changing from 1260 nm to 1340 nm.}
\label{fig: images_2}
\end{figure}
To further investigate the broadband behavior of the metalens, we repeated the measurement of the point spread function for a wavelength range centered around $\lambda$ = 1300 nm. Results are reported in Fig. \ref{fig: images_2} for normal and tilted illumination at 23° and 43°. Due to chromatic aberration, the image plane moved about 17 \mic along the z direction compared to the images of Fig. \ref{fig: images_1}(b), as detailed below. Despite being designed to operate at $\lambda$ = 1550 nm, a large and broadband field of view could be obtained also in this case, with well defined and undistorted focal spots for the different illumination angles and wavelengths between 1260 nm and 1340 nm.

\begin{figure}[t]
\centering\includegraphics[width=0.98\textwidth]{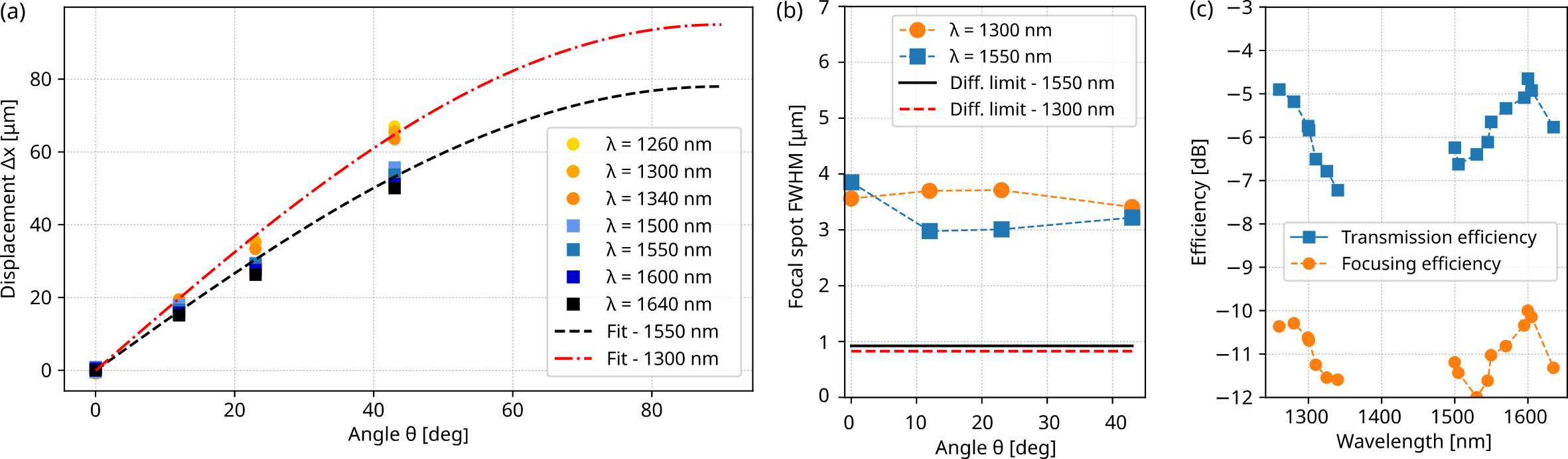}
\caption{Broadband performance of the metalens. (a) Displacement of the focal spot $\Delta x$ in the x direction as a function of the incident angle $\theta$ for a few sampled wavelengths. Transverse chromatic aberration is smaller than the spot size within both wavelength ranges around 1300 nm and 1550 nm. (b) Average full width at half maximum of the focal spot in the two wavelength ranges as a function of $\theta$. The diffraction limit is reported as well for reference. (c) Transmission and focusing efficiencies as a function of the wavelength. All measurements are done at fixed image planes for the two wavelength ranges.}
\label{fig: analysis}
\end{figure}

Figure \ref{fig: analysis} presents a more detailed analysis of the performance of the metalens. In particular, we quantified the effect of longitudinal and transverse chromatic aberration which cause the focal spot to move along the z and x axes, respectively, when wavelength is varied. While for lenses with a small field of view transverse chromatic aberration can often be overlooked, it becomes here a relevant parameter to define the off-axis behavior of the lens \cite{yang_design_2021}. The measured displacement $\Delta x$ along the x axis as a function of the illumination angle is shown in Figure \ref{fig: analysis}(a) for sampled wavelengths across both the 1500 nm - 1640 nm (blue squares) and 1260 nm - 1340 nm ranges (orange dots). As already mentioned, the 43° limit is due to our measurement setup and not to the metalens design. Full $\pm$90° field of view has already been demonstrated using quadratic metalenses \cite{martins_metalenses_2020}. The displacement is perfectly fit by Eq. \eqref{eq: transversal_shift} using a focal length $f$ = 79 \mic at $\lambda$ = 1550 nm (as expected from the design) and $f$ = 96 \mic at $\lambda$ = 1300 nm.

From the displacement measurements it is possible to estimate the effect of the longitudinal chromatic aberration within the two considered wavelength ranges. The maximum longitudinal chromatic shift from the mean focal length along the z axis is limited to 4 \mic for the 1550 nm range (5\% relative focal shift over a 140-nm bandwidth) and 3 \mic for the 1300 nm range (3\% relative focal shift over a 100-nm bandwidth). This confirms that the relatively small chromatic focal shifts combined with the large depth of focus enables the fixed image plane to fall within the focal tolerance of the lens throughout both spectral ranges, as previously shown in Figs. \ref{fig: images_1} and \ref{fig: images_2}. Regarding transverse chromatic aberration, the deviation of the displacement $\Delta x$ of the spot on the image plane caused by a variation of the wavelength for $\theta$ = 43° is about 3 \mic and 2 \mic in the measured ranges around $\lambda$ = 1550 nm and $\lambda$ = 1300 nm, respectively. This wavelength-dependent transverse shift becomes more pronounced as the illumination angle grows. However, at $\theta$ = 90° the maximum longitudinal focal shift also corresponds to the maximum deviation of the transverse displacement $\Delta x$, which is hence limited to 4 \mic and 3 \mic in the two ranges, respectively. This shift is approximately equal to the full width at half-maximum (FWHM) of the point spread function, measured as about 3.5 \mic, which makes the behavior of the lens close to achromatic also in the transverse direction. Smaller deviations can be obviously obtained either reducing the bandwidth or the field of view of the lens.

The angular dependence of the FWHM of the point spread function is shown in Fig. \ref{fig: analysis}(b). The average values across the two considered wavelength ranges are reported. As already mentioned, the quadratic phase profile introduces spherical aberration. As a result, the metalens does not reach the diffraction limit, which is reported for reference in the figure, and has a FWHM $\simeq$ 2$\lambda$ (the measurements of the FWHM have an uncertainty of about 1 \mic). This resolution is equivalent to that achieved by metalenses free of spherical aberration (such as hyperbolic ones) with a numerical aperture of about 0.3. Furthermore, at normal incidence, diffraction orders become evanescent at a distance from the lens axis larger than the focal distance and do not contribute to the focal spot, reducing the lens effective numerical aperture to about 0.71. It should also be noted, however, that a nominal numerical aperture larger than 0.8 is still essential to achieve an arbitrarily large field of view \cite{pu_nanoapertures_2017, martins_metalenses_2020}.  No significant differences can be observed between $\lambda$ = 1300 nm and $\lambda$ = 1550 nm and, importantly, the FWHM remains constant for increasing angle of incidence, which ensures the lens maintains good focusing properties also at tilted illumination. This is in marked contrast with hyperbolic metalenses that achieve diffraction-limited resolution but whose point spread function becomes heavily distorted even with small angles of incidence, making them unsuited for wide field of view operation.
Lastly, we quantified the wavelength dependence of the transmission and focusing efficiency of the lens, reported in Fig. \ref{fig: analysis}(c) at normal incidence ($\theta$ = 0°). The two efficiencies were determined by measuring the transmitted power in the area of the metalens and the transmitted power confined at the focal spot, respectively, normalized by the total power incident on the backside of the sample. Measurements were conducted at the same fixed image planes used in Fig. \ref{fig: images_1}(b) and \ref{fig: images_2}. The rather low transmission efficiency is due to the reduced effective aperture of the metalens while spherical aberration determines a further drop in focusing efficiency. It is interesting, however, how both transmission and focusing efficiencies exhibit a weak dependence on wavelength, with less than 2 dB fluctuations over the entire bandwidth between 1260 nm and 1640 nm. This is achieved despite the fact the meta-atoms used to realize the lens have been selected according to their transmission at $\lambda$ = 1550 nm, as reported in Fig. \ref{fig: design}(d). Additionally, the focusing efficiency of the lens with respect to the transmitted power fluctuates between 22\% (-6.6 dB) and 36\% (-4.4 dB) over the 380-nm bandwidth.

\section{Conclusion}
In summary, we have experimentally demonstrated that a single-layer metalens implementing a quadratic phase profile can simultaneously achieve a wide field of view and broadband achromatic behaviour. We realized a silicon-based metalens for the near-infrared and we characterized its performance at two wavelength ranges around $\lambda$ = 1300 nm and $\lambda$ = 1550 nm, respectively. Under tilted illumination, the focal spot of the lens remained free of off-axis aberrations and simply exhibited a transverse shift consistently throughout both spectral ranges, potentially reaching an arbitrarily large field of view. The FWHM of the focal spot remained almost constant at about 3.5 \mic independently of the wavelength and illumination angle. Longitudinal chromatic aberration was compensated by the relatively large depth of focus of the lens. The main limit to achromatic behavior was instead determined by transverse chromatic aberration which caused the position of the focal spot in the transverse direction for a given illumination angle to change with wavelength. However, even considering the largest possible field of view, transverse chromatic shift remained at most as large as the FWHM of the focal spot over a bandwidth up to 140 nm. Further extension of the operational bandwidth of the lens may be achieved by compensating for chromatic aberrations at design level. Beside the great potentialities as ultra-wide lenses for imaging applications \cite{martins_metalenses_2020}, we believe the use of quadratic metalenses may prove particularly advantageous in the context of beam steering, for example in combination with focal plane switch arrays \cite{rogers_universal_2021}. In this case, diffraction-limited resolution is not required and the larger focal spot of quadratic metalenses does not represent a limitation for the performance of the system. On the contrary, their achromatic behaviour and arbitrarily large field of view would offer an unprecedented route toward broadband photonic systems, paving the way to miniaturized devices for multispectral lidars \cite{li_spectral_2021} and free-space optical communications based on wavelength division multiplexing.

\section{Acknowledgments}
The fabrication of the device was performed at the Plateforme de Micro-Nano-Technologie/C2N, which is partially funded by the Conseil General de l’Essonne. This work was partly supported by the French RENATECH network and by the Agence Nationale de la Recherche (ANR) under the Tremplin – ERC project ANR-22-ERCS-0007-01.

%%%%%%%%%%%%%%%%%%%%%%% References %%%%%%%%%%%%%%%%%%%%%%%%%
\bibliography{references}

\end{document}